\documentclass[journal]{IEEEtran}
\usepackage{amsmath}
\usepackage{amssymb}
\usepackage{xcolor}
\usepackage{float}
\usepackage{lipsum}
\usepackage{framed}
\usepackage{lipsum}
\usepackage{bm}
\usepackage{cuted}
\usepackage{lipsum}
\usepackage{graphicx}
\usepackage{stfloats}
\usepackage{color}
\usepackage{epstopdf}
\begin{document}
\title{Performance Analysis of an Interference-Limited RIS-Aided Network}
\author{Liang Yang, Yin Yang, Daniel Benevides da Costa, and Imene Trigui
\thanks{L. Yang and Y. Yang are with the College of Information Science and Engineering, Hunan University, Changsha
410082, China, (e-mail:liangy@hnu.edu.cn, yy19971417@163.com).}
\thanks{D. B. da Costa is with the Department of Computer Engineering, Federal
University of Cear\'{a}, Sobral, CE, Brazil (email: danielbcosta@ieee.org).}
\thanks{I. Trigui is with the University of Quebec, Montreal, QC, Canada (e-mail:
trigui.imene17@gmail.com).}}
\maketitle

\begin{abstract}
In this work, the performance of reconfigurable intelligent surface (RIS)-aided communication systems corrupted by the co-channel interference (CCI) at the destination is investigated. Assuming Rayleigh fading and equal-power CCI, we present the analysis for the outage probability (OP), average bit error rate (BER), and ergodic capacity. In addition, an asymptotic outage analysis is carried in order to obtain further insights. Our analysis shows that the number of reflecting elements as well as the number of interferers have a great impact on the overall system performance.
\end{abstract}

\section{Introduction}
A new disruptive technology, called reconfigurable intelligent surface (RIS), has recently emerged in the research community. The RIS is an artificial surface composed of electromagnetic (EM) materials that is electronically controlled by low-cost electronic devices. Owing to its unique functions, RIS can efficiently customize the wireless environment, thereby maximizing the signal quality at the receiver \cite{1}. Compared with other competing technologies, it is worth noting that RIS does not require encoding and decoding operations during the signal transmission. In addition, it does not create new waves and can change the shape of the wireless signal through soft programming. 

Recently, several works have investigated the incorporation of RISs in wireless systems. Specifically, the authors in \cite{2} studied the joint optimization of transmitting source and RIS beams in multiple-input single-output (MISO) communication systems. In \cite{3}, the authors studied the optimization of joint transmission and reflected beamforming in RIS-assisted multiuser systems. In \cite{4}, the application of RISs in downlink multiuser communications was examined, while \cite{5} provided an accurate analytical results for the coverage, probability of signal-to-noise ratio (SNR) gain, and delay outage rate of RIS-assisted wireless communication systems. Moreover, the analysis in \cite{6} showed that using RISs has a positive effect on improving the secrecy performance of wireless systems. A performance analysis for the application of RISs in a dual-hop free-space optical (FSO)/radio-frequency (RF) system was carried out in \cite{7}. Finally, \cite{8} analyzed the performance of RIS-aided unmanned aerial vehicle (UAV) communication relaying systems. However, for all we know, the impact of co-channel interference (CCI) in RIS-aided communication system remains to be investigated in the literature yet, which motivates this new work.


In this paper, the performance of RIS-assisted interference-limited communication systems with CCI at the destination is studied. To this end, we first derive the statistical distribution of the effective signal-to-interference ratio (SIR) expression and, based on it, exact analysis for the outage probability (OP), average bit error rate (BER), and average channel capacity are presented. To clearly show the effects of the interference and power levels on the system performance, an asymptotic outage expression is also derived. Our results reveal that the number of reflecting meta-surfaces as well as the number of interferers have a significant impact on the overall system performance.

\section{System and Channel Models}
\begin{figure}
\centering
\includegraphics[height=5cm,width=6cm]{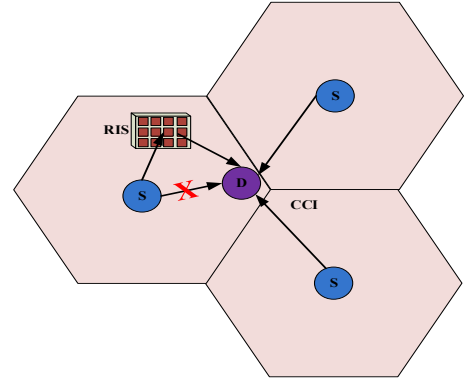}
\caption{System model.}
\end{figure}
Our considered system model can be found in Fig. 1, where a single-antenna source (S) intends to communicate with a single-antenna destination (D) via an
RIS formed by $N$ reflecting elements. The terminal D is affected by $L$ independent equal-power interferers. Such a scenario finds applicability in cases where the RIS and S are located at the center of a cellular network, whereas D is located at its border. We assume that direct link transmission between S and D does not exist due to obstacles. Moreover, it is assumed that the considered system operates in an interference-limited regime so that the effect of noise on the overall performance can be negligible. Similar to \cite{1}, we assume that the RIS has the ability to obtain the full channel state information (CSI). Thus, the resulting signal at D can be written as
\begin{equation}
y=\sqrt{P_{s}}\left [ \sum_{i=1}^{N}h_{i}\rho _{i}g_{i} \right ]x+\sum_{l=1}^{L}\sqrt{P_{I}}h_{l}x_{l},
\end{equation}
where $ P_{s}$ denotes the average transmitted power per symbol at the S, $x$ is the source signal, $ P_{I}$ denotes the transmit power of the interferers, and $h_{l}$ and $x_{l}$ are, respectively, the channel coefficient suffering from the Rayleigh fading and unit energy
signal for the $l\rm th$ interferer. In (1), $\rho_{i}=\omega_{i}(\phi_{i})e^{j\phi_{i}}$ denotes the reflection coefficient generated by the $i$th reflecting element of the RIS, with $\omega_{i}(\phi_{i})=1$ corresponding to ideal phase
shifts ($i = 1, 2, \ldots, N$). In addition, $h_{i}$ and $g_{i}$ refer to the channel coefficients of the S-RIS and RIS-D links, where $h_{i}=\alpha_{i}e^{-j\theta_{i}}$ and $g_{i}=\beta_{i}e^{-j\varphi_{i}}$, with $\theta_{i}$ and $\varphi_{i}$ denoting the respective phases of
the fading channel coefficients, and $\alpha_{i}$ and $\beta_{i}$ denote the respective channels' amplitudes, which are independent Rayleigh random variables (RVs).  

From \cite{1}, in order to achieve the maximum SIR, RIS can fully eliminate the phase shifts by supposing $\phi_{i}=\theta_{i}+\varphi_{i}$. Therefore, the maximum SIR is
\begin{equation}
\gamma _{\rm SIR}=\frac{P_{s}\left | \sum_{i=1}^{N} \alpha _{i}\beta _{i}\right |^{2}}{P_{I} \sum_{l=1}^{L}\left | h_{l}\right |^{2}}.
\end{equation}

Let $Y=P_{s}\left | \sum_{i=1}^{N} \alpha _{i}\beta _{i}\right |^{2}=P_{s}Z^{2}$. From \cite{9}, the probability density function (PDF) of $Y$ can be modeled by a squared $K_{G}$ distribution, i.e.,
\begin{equation}
f_{Y}(y)=\frac{2A^{k+m}}{\Gamma (k)\Gamma (m)P_{s}^{\frac{k+m}{2}}}y^{\left ( \frac{k+m}{2}-1\right )}K_{k-m}\left ( 2A\sqrt{\frac{y}{P_{s}}} \right ),
\end{equation}
where $A=\sqrt{km/\Omega}$, $\Omega\overset{\triangle }{=}E(Z^{2})$ is the mean power,
 $k$ and $m$ are the shaping parameters, $K_{\nu}(\cdot)$ is
the modified Bessel function of the second kind with zero
order \cite{10}, and $\Gamma(\cdot)$ denotes the gamma function \cite{10}. Now, let $X=P_{I} \sum_{l=1}^{L}\left | h_{l}\right |^{2}$. From \cite{11}, we know that $X$ is a chi-squared RV with $2L$ degrees of
freedom and whose PDF is given by
\begin{equation}
f_{X}(x)=\frac{1}{P_{I}^{L}\Gamma(L)}x^{L-1}e^{-\frac{x}{P_{I}}}.
\end{equation}

Using (3) and (4), the
PDF of $\gamma_{\rm SIR}$ can be written as
\begin{align}
f_{\gamma _{\rm SIR}}(\gamma )=&\int_{0}^{\infty }xf_{Y}(\gamma x)f_{X}(x)dx\nonumber\\
=&\frac{A^{k+m}}{\Gamma (k)\Gamma (m)P_{s}^{\frac{k+m}{2}}\Gamma (L)}\gamma ^{\frac{k+m}{2}-1}P_{I}^{\frac{k+m}{2}}\nonumber\\
\times&G_{1,2}^{2,1}\left[\left.\frac{A^2\gamma P_{I}  }{P_{s}}\right| \begin{matrix}1-L-\frac{k+m}{2}\\\frac{k-m}{2},\frac{m-k}{2}\end{matrix}\right],\nonumber
\tag{5}
\end{align}
where $G_{c,d}^{a,b}\left [ \cdot  \right ]$ denotes the Meijer G-function defined in \cite{10}.
Furthermore, the cumulative
distribution function (CDF) of $\gamma_{\rm SIR}$ can be shown to be given by
\begin{align}
F_{\gamma _{\rm SIR}}(\gamma )=&\frac{A^{k+m}}{\Gamma (k)\Gamma (m)P_{s}^{\frac{k+m}{2}}\Gamma (L)}\gamma ^{\frac{k+m}{2}}P_{I}^{\frac{k+m}{2}}\nonumber\\
\times&G_{2,3}^{2,2}\left[\left.\frac{A^2\gamma P_{I}  }{P_{s}}\right| \begin{matrix}1-\frac{k+m}{2},1-L-\frac{k+m}{2}\\\frac{k-m}{2},\frac{m-k}{2},-\frac{k+m}{2}\end{matrix}\right].\nonumber
\tag{6}
\end{align}
\section{Performance Analysis}
In this section, we investigate the OP, average BER, and average channel capacity of our considered model under the impact of equal-power CCI. In addition, we carry out an asymptotic analysis by assuming high $P_{s}$ and $P_{I}$ to get further insights, which may be helpful for the system design.
\subsection{Outage Probability}
\emph{1. Exact Analysis}

From \cite{12}, the definition of OP is the probability that
the end-to-end SIR $\gamma_{\rm SIR}$ is lower than a preset threshold
$\gamma_{\rm th}$, which is mathematically written as $P_{out}=\Pr(\gamma_{\rm SIR}<\gamma_{\rm th})$.

By inserting $\gamma=\gamma_{\rm th}$ into (6), the OP
can be written as
\begin{equation}
P_{out}=F_{\gamma _{\rm SIR}}(\gamma_{\rm th} ).
\tag{7}
\end{equation}
\emph{2. Asymptotic Analysis}

According to \cite{13}, note that the PDF of $Y$ can be formulated as $f_{Y}(y)=\Phi y^{\iota}+o(y)$, where $\Phi$ is
a positive constant, $\iota$ represents the quantization of the
smoothing order of $f_{Y}(y)$ at the origin, and $o(y)$ denotes the higher order terms. Therefore, one can obtain an asymptotic PDF expression for (3) as
\begin{equation}
f_{Y}(y)\simeq\frac{2A^{k+m}}{\Gamma (k)\Gamma (m)P_{s}^{\frac{k+m}{2}}}y^{\left ( \frac{k+m}{2}-1\right )},
\tag{8}
\end{equation}
and its corresponding asymptotic PDF expression as
\begin{equation}
f_{\gamma _{\rm SIR}}(\gamma )\simeq\frac{2A^{k+m}\Gamma (\frac{k+m}{2}+L)P_{I}^{\frac{k+m}{2}}}{\Gamma (k)\Gamma (m)P_{s}^{\frac{k+m}{2}}\Gamma (L)}\gamma ^{\left ( \frac{k+m}{2}-1\right )}.
\tag{9}
\end{equation}

Therefore, the asymptotic outage expression can be derived as
\begin{equation}
P_{out}\simeq\frac{4A^{k+m}\Gamma (\frac{k+m}{2}+L)P_{I}^{\frac{k+m}{2}}}{\Gamma (k)\Gamma (m)P_{s}^{\frac{k+m}{2}}\Gamma (L)(k+m)}\gamma _{\rm th}^{\frac{k+m}{2}}.
\tag{10}
\end{equation}

The above expression clearly shows that the diversity order is $\frac{k+m}{2}$, where $k$ and $m$ denote the pair of conjugate complex numbers, and are determined by $N$. As expected, by increasing the number of interferers $L$ or the interference power $P_{I}$ will lead to an increase in the OP, thereby reducing the system performance. In addition, by increasing the power of S will make the system performance better.

\subsection{Average BER}
The performance metric BER is a main indicator to evaluate the accuracy of the signal transmission.
From \cite{14}, the average BER can be expressed as
\begin{equation}
P_{e}=\frac{q^{p}}{2\Gamma (p)}\int_{0}^{\infty }e^{-q\gamma }\gamma ^{p-1}F_{\gamma _{\rm SIR}}(\gamma )d\gamma,
\tag{11}
\end{equation}
where different values of $p$ and $q$ correspond to various modulation schemes. For instance, differential phase shift keying (DPSK) ($q=1$ and $p=1$), binary phase shift keying (BPSK) ($q=\frac{1}{2}$ and $p=1$), and binary frequency shift keying (BFSK) ($q=\frac{1}{2}$ and $p=\frac{1}{2}$). In this work, we consider the DPSK and the BPSK schemes.

From (6) and (11),  and using [15, Eq.(07.34.21.0088.01)], the average BER can
be evaluated as
\begin{align}
P_{e}=&\frac{q^{-\frac{k+m}{2}}A^{k+m}P_{I}^{\frac{k+m}{2}}}{2\Gamma (p)\Gamma (k)\Gamma (m)P_{s}^{\frac{k+m}{2}}\Gamma (L)}\nonumber\\
\times&G_{3,3}^{2,3}\left[\left.\frac{A^2P_{I}  }{P_{s}q}\right| \begin{matrix}1-p-\frac{k+m}{2},1-\frac{k+m}{2},1-L-\frac{k+m}{2}\\\frac{k-m}{2},\frac{m-k}{2},-\frac{k+m}{2}\end{matrix}\right].\nonumber
\tag{12}
\end{align}
\subsection{Average Channel Capacity}
From \cite{16}, we have
\begin{equation}
C=\frac{1}{\rm ln(2)}\int_{0}^{\infty }\rm ln(1+\gamma )\emph f_{\gamma _{SIR}}(\gamma )\emph d\gamma.
\tag{13}
\end{equation}

For a simple calculation of the capacity, one can apply a useful identity in [15, Eq.(01.04.26.0003.01)], i.e., $\rm ln(1+\gamma )=\emph G_{2,2}^{1,2}\left[\left.\gamma \right| \begin{matrix}1,1\\1,0\end{matrix}\right]$.
By combining [15, Eq.(07.34.21.0011.01)]
together with (5) and (13), the average capacity can
be evaluated as
\begin{align}
C=&\frac{A^{k+m}}{\Gamma (k)\Gamma (m)P_{s}^{\frac{k+m}{2}}\Gamma (L)\rm ln(2)}\gamma ^{\frac{k+m}{2}}P_{I}^{\frac{k+m}{2}}\nonumber\\
\times&G_{3,4}^{4,2}\left[\left.\frac{A^2 P_{I}  }{P_{s}}\right| \begin{matrix}1-L-\frac{k+m}{2},-\frac{k+m}{2},1-\frac{k+m}{2}\\\frac{k-m}{2},\frac{m-k}{2},-\frac{k+m}{2},-\frac{k+m}{2}\end{matrix}\right].\nonumber
\tag{14}
\end{align}
\section{Numerical Results and Discussions}
In this section, we select some numerical examples to verify the impact of the main system parameters on the system performance.
Our analysis is confirmed by Monte Carlo simulations, which generates $10^5$ simulation points. In the next figures, the SIR threshold is set to $\gamma_{\rm th}=20\, \rm dB$.
\begin{figure}
\centering
\includegraphics[height=6cm,width=8cm]{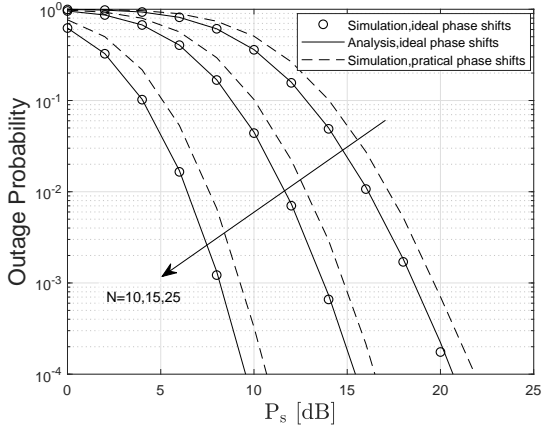}
\caption{Outage probability versus $P_{s}$ for different number of $N$.}
\end{figure}

In Fig. 2, we present the OP performance of our considered system with
various values of $N$, and by setting the number of interferers $L =4$, with all interferers having the same
power, i.e., $ P_{I}=1 \, \rm dB$. From this figure, one can find that the analytical results match the simulation results perfectly. In addition, the simulation results of the ideal and practical
phase shifts are given to observe the performance hit. As mentioned in [3], one can use $\omega_{i}(\phi_{i})=(1-\varpi_{\rm min})\left(\frac{\rm sin(\phi_{\emph i}-\kappa)+1}{2}\right )^{\varsigma} + \varpi_{\rm min}$ to represent the performance loss when considering the practical phase shifts, where $\varpi_{\rm min}$ stands for the minimum amplitude, $\kappa$ represents the horizontal distance between $\frac{\pi}{2}$
and $\varpi_{\rm min}$, and $\varsigma$ denotes the steepness of function curve.
We consider the setup $\varpi_{\rm min}=0.8$, $\kappa=0.43\pi$, and  $\varsigma=1.6$.
One can clearly see that there is a certain performance difference between the practical and ideal cases.
Finally, it can be observed that applying large values of $N$ can result in a better system performance.
\begin{figure}
\centering
\includegraphics[height=6cm,width=8cm]{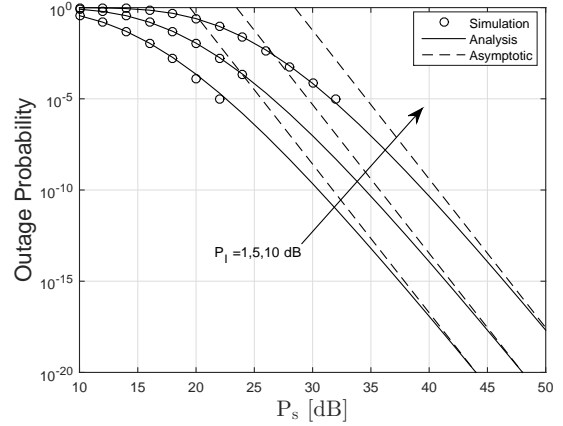}
\caption{Outage probability versus $P_{s}$ for different values of $P_{I}$.}
\end{figure}

In Fig. 3, we plot the OP curves for various values of $P_{I}$ and by setting $L = 4$ and $N=10$. As can be seen, increasing $P_{I}$ leads to a deterioration of the OP performance. In Fig. 4, the OP is plotted for various values of $L$ and assuming $P_{I}=1 \, \rm dB$ and $N=10$. The results show that increasing the amount of interference reduces the system performance. From both Figs. 3 and 4, one can be clearly seen that all curves have the same slopes, which means that the existence of interference does not affect the diversity order of the considered system, corroborating the fact that the diversity order depends on $N$. Also, at high $P_{s}$, the asymptotic results are close to the exact values.

In Fig. 5, we plot the BER curves for both DPSK and BPSK schemes when $ P_{I}=1 \, \rm dB$, $L = 4$ and $N = 10$. One can clearly observe that the BPSK scheme has a better system performance.

Finally, Fig. 6 depicts the average capacity of our considered system for various values
of $N$ and assuming $L = 8$ and $P_{I}=1 \, \rm dB$. Again, it can be clearly seen that the analytical results match perfectly the simulation ones. Moreover, the system
performance improves as $N$ increases.
\begin{figure}
\centering
\includegraphics[height=6cm,width=8cm]{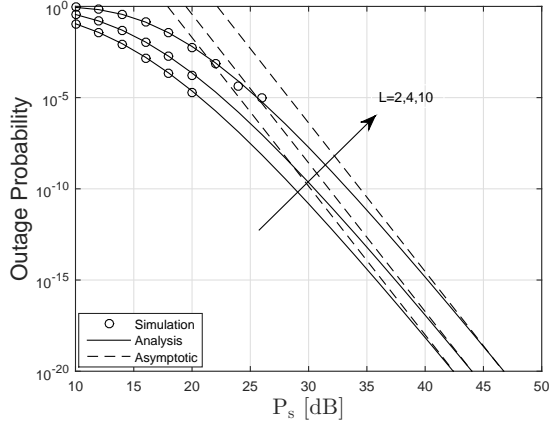}
\caption{Outage probability versus $P_{s}$ for different numbers of co-channel interferers.}
\end{figure}
\begin{figure}
\centering
\includegraphics[height=6cm,width=8cm]{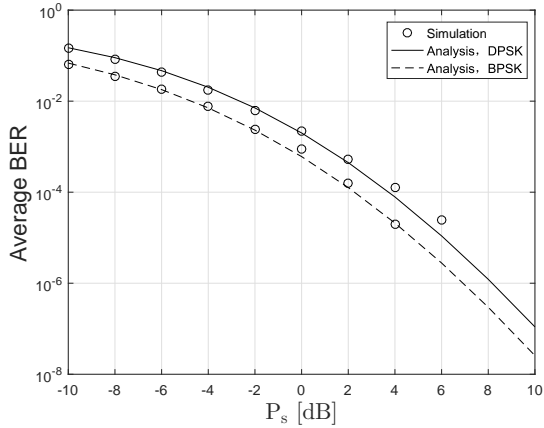}
\caption{Average BER versus $P_{s}$ for different modulation schemes.}
\end{figure}
\begin{figure}
\centering
\includegraphics[height=6cm,width=8cm]{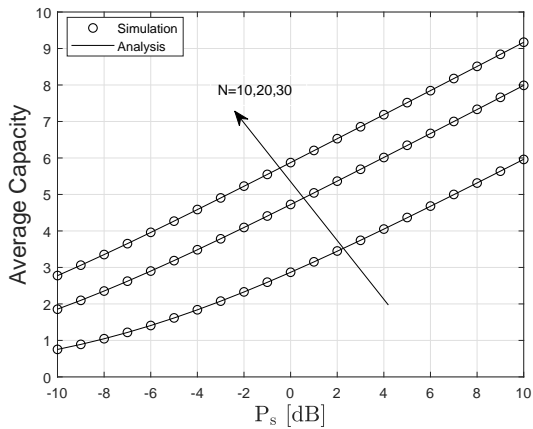}
\caption{Average capacity versus $P_{s}$ for different number of $N$.}
\end{figure}
\section{Conclusion}
In this paper, we presented a performance analysis for RIS-assisted communication systems with equal power CCI. We first performed an accurate analysis of the OP, average BER, and average capacity. The results showed that our analysis results match the simulation ones perfectly. In order to clearly study the effect of parameters on the system performance, we derived an asymptotically closed-expression for the OP. The results revealed that the achievable diversity order equals to $\frac{k+m}{2}$, and the number of interferers has a great influence on the system performance.

\end{document}